# Vacancies and dopants in two-dimensional tin monoxide: An *ab initio* study

Devesh R. Kripalani,[1, 2] Ping-Ping Sun,[1] Pamela Lin,[2] Ming Xue,[2] and Kun Zhou[1, 3, *]

[1] *School of Mechanical and Aerospace Engineering, Nanyang Technological University, 50 Nanyang Avenue, Singapore 639798, Singapore*

[2] *Infineon Technologies Asia Pacific Pte Ltd, 8 Kallang Sector, Singapore 349282, Singapore*

[3] *Environmental Process Modelling Centre, Nanyang Environment and Water Research Institute, Nanyang Technological University, 1 Cleantech Loop, Singapore 637141, Singapore*

Corresponding author(s):

* kzhou@ntu.edu.sg

## Abstract

Layered tin monoxide (SnO) offers an exciting two-dimensional (2D) semiconducting system with great technological potential for next-generation electronics and photocatalytic applications. Using a combination of first-principles simulations and strain field analysis, this study investigates the structural dynamics of oxygen (O) vacancies in monolayer SnO and their functionalization by complementary lightweight dopants, namely C, Si, N, P, S, F, Cl, H and $H_2$. Our results show that O vacancies are the dominant native defect under Sn-rich growth conditions with active diffusion characteristics that are comparable to that of graphene vacancies. Depending on the choice of substitutional species and its concentration within the material, significant opportunities are revealed in the doped-SnO system for facilitating *n*/*p*-type tendencies, work function reduction, and metallization of the monolayer. N and F dopants are found to demonstrate superior mechanical compatibility with the host lattice, with F being especially likely to take part in substitution and lead to degenerately doped phases with high open-air stability. The findings reported here suggest that post-growth filling of O vacancies in Sn-rich conditions presents a viable channel for doping 2D tin monoxide, opening up new avenues in harnessing defect-engineered SnO nanostructures for emergent technologies.

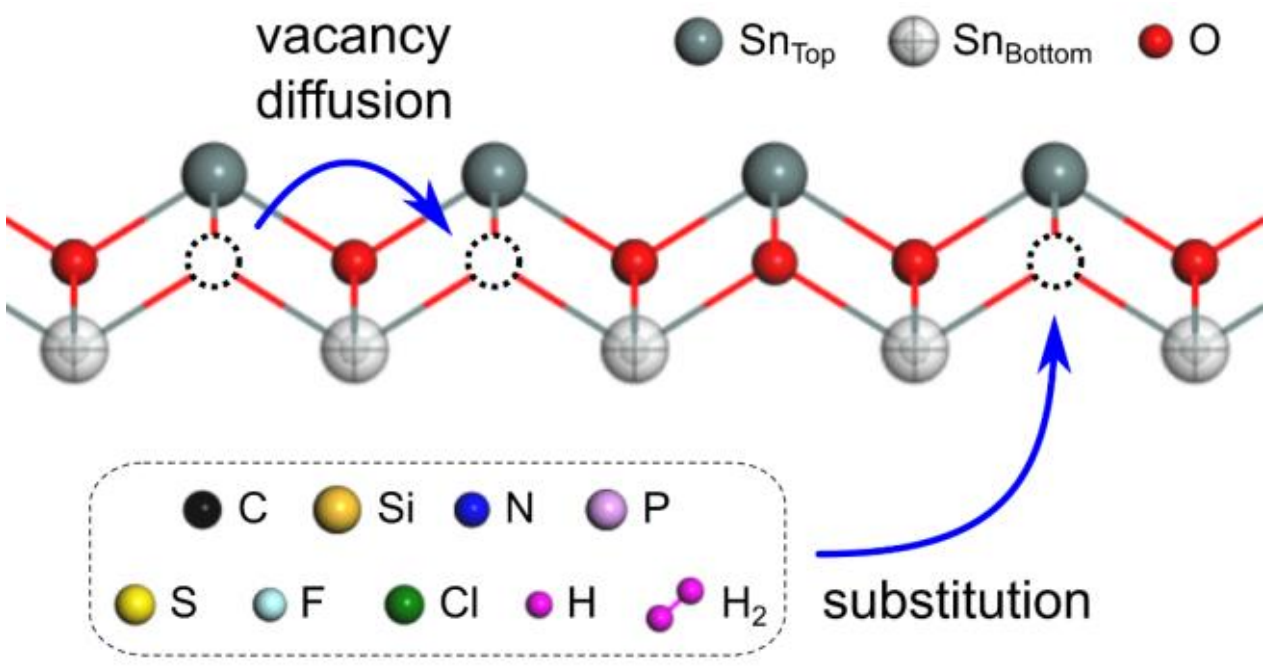

**Graphical abstract**

## 1. Introduction

Layered metal oxides (LMOs) and their nanoscale derivatives represent an emerging class of compounds in the field of two-dimensional (2D) materials beyond graphene [1]. LMOs are usually air-stable by nature and are composed of non-toxic, earth-abundant elements, making them environmentally desirable and cost-effective for large-scale production [2-4]. Held together by weak van der Waals (vdW) interactions, LMO nanosheets can be easily isolated using conventional exfoliation techniques as demonstrated successfully in layered oxides of Ti, Nb, Mn and Co [4-6]. In the monolayer, a number of such oxides have been identified as wide-bandgap semiconductors ($E_g$ > 2 eV) with high carrier mobility and strong optical absorption in the UV-vis range [3]. Today, LMOs continue to gain broad research interest due to their versatility for a myriad of applications, including thin-film transistors (TFTs) [7, 8], capacitive energy storage [9], optoelectronics [10, 11], photocatalytic water splitting [12], and even molecular sensing [13].

Recently, layered tin monoxide (SnO) has attracted considerable attention for its bipolar semiconducting features [14, 15]. Structurally, SnO adopts a tetragonal (litharge-type) crystal lattice ($a = b = 3.80$ Å, $c = 4.84$ Å) [16, 17], with each layer consisting of Sn-O-Sn atomic planes oriented along the [001] crystallographic direction (see Fig. 1(a)). Despite being intrinsically *p*-type due to the presence of Sn vacancies as the primary native defect [18], doping with Sb has been found to enable *n*-type conductivity [15]. Moreover, under suitable experimental conditions, SnO has shown great potential as active channel layers in TFT-based devices [19-21], as anode materials in Li-/Na-ion batteries [22, 23] and as efficient photoelectrodes for water splitting [24]. Advances in synthesis methods like liquid phase exfoliation [25] and pulsed laser deposition [19] have also offered improved precision and control over the thickness of fabricated SnO nanosheets, providing a means to regulate their optical and electronic properties such as the bandgap, which is known to range from 0.7 eV (indirect) in bulk to as high as ~4 eV in the monolayer. However, as research efforts have traditionally focused on thin-film (polycrystalline/bulk) structures of SnO, little is understood about its behavior in the 2D limit, particularly when the thickness is reduced to the order of several atoms (i.e., in the sub-nm regime).

The chemistry of defects in 2D materials are typically of fundamental interest given their ubiquitous impact on optoelectronic and magnetic properties [26-29]. Attempts to deliberately engineer defect states, for example via careful control of processing conditions and/or introducing functional dopants, have been extensively documented in literature and can be used to enhance or even repurpose low-dimensional materials for desired applications [30-32]. So far, strategies for tuning the magnetic character of 2D SnO through surface adsorption and transition metal doping of Sn vacancies have been theoretically reported, suggesting possibilities for its use in spintronic devices [33, 34]. On the other hand, the influence of O vacancies and their functionalization by dopants remains an underexplored issue deserving of further attention, especially since such defects can be active in Sn-rich growth conditions [33] as well as in the monolayer where they are directly exposed to the external environment.

In the present study, the physical chemistry of O vacancies and various O-site dopant candidates (C, Si, N, P, S, F, Cl, H, $H_2$) in monolayer SnO are systematically investigated based on first-principles calculations. All dopants selected in this study are either smaller than or of comparable size to oxygen and are hence expected *a priori* to be mechanically compatible with the host lattice. Furthermore, for completeness, the effect of dopant concentration (i.e., isolated vs. degenerate doping) is also considered in this work. Insights into the structural energetics and diffusion kinetics of O vacancies in 2D tin monoxide are first discussed in Sec. 3.1., while new opportunities for tailoring its electronic characteristics (e.g., *n*/*p*-type conductivity, work function and band structure) through targeted substitutional doping are disclosed in Sec. 3.2. and Sec. 3.3. In Sec. 4, the main findings of this research are summarized, and future works are recommended.

## 2. Computational method

Spin-polarized first-principles calculations are performed based on the Perdew-Burke-Ernzerhof (PBE) [35] exchange-correlation functional within the framework of density functional theory (DFT) [36, 37] using the Vienna *ab initio* simulation package (VASP) [38]. The DFT-D2 method of Grimme [39] is applied to describe the vdW interactions in layered tin monoxide. This approach, which we refer to as PBE+D2, has been previously implemented in our earlier study on ultrathin SnO films [40] and is verified to be able to reproduce the structure of bulk SnO with good agreement to experiment. At the same time, it can provide reliable results (in a qualitative sense) for the electronic properties, namely, the band structure and work function. Spin-orbit coupling is confirmed to have negligible effects on the electronic band structure of monolayer SnO (see Fig. S1 in the supplementary material (SM)) and have not been included in our calculations. Projector augmented-wave (PAW) pseudopotentials are used to describe ion-electron interactions, with $4d^{10}5s^{2}5p^{2}$, $2s^{2}2p^{4}$, $2s^{2}2p^{2}$, $3s^{2}3p^{2}$, $2s^{2}2p^{3}$, $3s^{2}3p^{3}$, $3s^{2}3p^{4}$, $2s^{2}2p^{5}$, $3s^{2}3p^{5}$, and $1s^{1}$ treated as valence electrons for Sn, O, C, Si, N, P, S, F, Cl, and H, respectively.

Isolated vacancies (V) and dopants (X) in monolayer SnO are simulated in a 5 × 5 × 1 supercell. In the context of substitutional doping at the O site, this corresponds to a $SnO_{(1-\delta)}X_{\delta}$ system with a dopant concentration $\delta = 0.02$. Doping in the degenerate limit $\delta = 0.5$ is addressed by using a unit cell model. A kinetic energy cutoff of 500 eV is selected for the plane wave basis set, and the Brillouin zone is sampled with a 2 × 2 × 1 (10 × 10 × 1) *k*-point mesh for the supercell (unit cell) using the Monkhorst-Pack method. The energy convergence criteria for electronic iterations is set at $10^{-6}$ eV and all structures are relaxed until the maximum force per atom is below 0.02 eV/Å. Free boundary conditions are prescribed in the normal (*z*) direction by introducing a vacuum separation layer of ~20 Å between adjacent slabs. The climbing image nudged elastic band (CINEB) [41] method is adopted to investigate the migration paths and diffusion barriers of vacancies in the system.

The formation energy $E_{\mathrm{f}}$ of a defect is given by,

$$E_{\mathrm{f}} = E_{\mathrm{defect}} - (E_{\mathrm{pristine}} - \mu_{\mathrm{host}} + \mu_{\mathrm{X}}) \tag{1}$$

where $E_{\mathrm{pristine}}$ ($E_{\mathrm{defect}}$) refers to the total energy of the pristine (defective) system, while $\mu_{\mathrm{host}}$ and $\mu_{\mathrm{X}}$ are the chemical potentials of the removed/replaced host atom and the substitutional dopant, respectively. The chemical potentials of the dopants are computed with reference to the diamond structure of bulk C and Si, and the diatomic molecules of N, P, S, F, Cl, and H.

Atomic chemical potentials of the dopants are also considered in this study; this case is denoted by $\mu_X$* and $E_f$* accordingly. The formation energy per formula unit (f.u.) of the SnO monolayer itself, $E_f^{SnO}$, is defined by,

$$E_{\mathrm{f}}^{\mathrm{SnO}} = \mu_{\mathrm{SnO}} - (\mu_{\mathrm{Sn}}^0 + \mu_{\mathrm{O}}^0) \quad (2)$$

where $\mu_{SnO}$ is the total energy per f.u. of monolayer SnO and $\mu_{Sn}^0$ ($\mu_O^0$) is the chemical potential of Sn (O) in its reference phase. As the choice of reference phase is arbitrary, we have taken them to be the standard states [42] of tin (bulk $\beta$-Sn, body-centered tetragonal (BCT)) and oxygen ($O_2$ molecule) herein. Therefore, we obtain $E_f^{SnO} = -2.435$ eV.

The value of $\mu_{host}$ in Eq. (1), and hence also that of $E_f$, is dependent on experimental growth conditions, which can range between two limiting scenarios: Sn-rich and O-rich conditions. Note that Sn-rich (O-rich) conditions are analogous to O-poor (Sn-poor) conditions as they are sometimes referred to in literature [33, 43]. Under Sn-rich conditions, the chemical potential of Sn is equal to that of its reference phase (i.e., $\mu_{Sn}^{\text{Sn-rich}} = \mu_{Sn}^0$). In thermodynamic equilibrium, one can assume that $\mu_{SnO} = \mu_{Sn} + \mu_O$, such that for Sn-rich conditions, the chemical potential of O is $\mu_O^{\text{Sn-rich}} = \mu_{SnO} - \mu_{Sn}^0$. With the aid of Eq. (2), the chemical potentials for Sn-rich conditions can then be expressed as,

$$\mu_{\mathrm{Sn}}^{\mathrm{Sn-rich}} = \mu_{\mathrm{Sn}}^0 \quad (3a)$$

$$\mu_{\mathrm{O}}^{\mathrm{Sn-rich}} = \mu_{\mathrm{O}}^0 + E_{\mathrm{f}}^{\mathrm{SnO}} \quad (3b)$$

In the same way, the chemical potentials for O-rich conditions can be derived as,

$$\mu_{\mathrm{Sn}}^{\mathrm{O-rich}} = \mu_{\mathrm{Sn}}^0 + E_{\mathrm{f}}^{\mathrm{SnO}} \quad (4a)$$

$$\mu_{\mathrm{O}}^{\mathrm{O-rich}} = \mu_{\mathrm{O}}^0 \quad (4b)$$

The relevant data of all the chemical species involved in this work are listed in Table S1 in the SM.

## 3. Results and discussion

### 3.1. Energetics and kinetics of oxygen vacancy defects

Prior to our study on the O-site substitutional system, the energetics and kinetics of the oxygen vacancy defect ($V_O$) in monolayer SnO is first characterized in detail. The optimized structural parameters of the pristine monolayer in its 2D-tetragonal (square) lattice, as shown in Fig. 1(a), are calculated to be $a = b = 3.83$ Å and $t = 2.39$ Å. Its internal stacking registry of Sn-O-Sn atomic planes along the normal ($z$) direction indicates that O vacancies are inherently confined within the monolayer and as such, their physicochemical properties will be largely moderated through screening by surface Sn atoms.

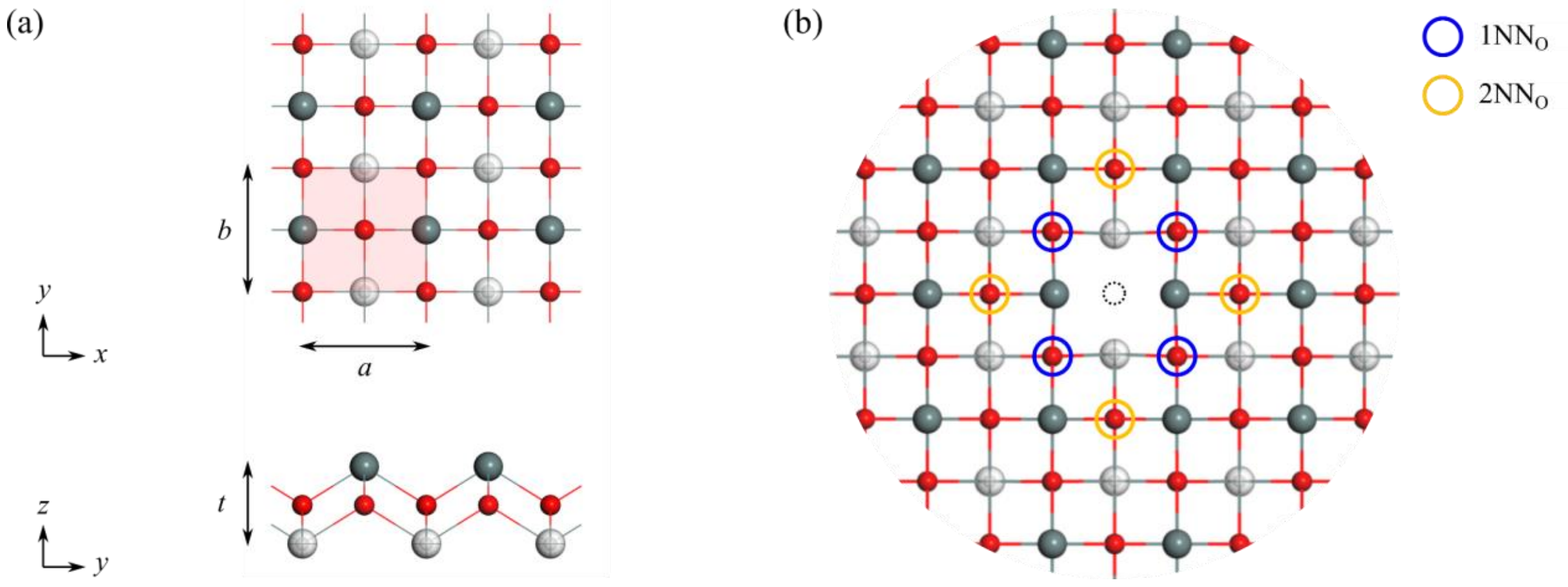

**Figure 1.** (a) 2D-tetragonal crystal structure of pristine monolayer SnO (O, red; top/bottom Sn, dark/light grey), with the unit cell (shaded pink) indicated in the top ($xy$) view. (b) Relaxed structural configuration (top view) of monolayer SnO containing an O vacancy defect. The defect site is denoted by a dashed circle, while its first- and second-nearest O neighbors, $1NN_O$ and $2NN_O$, are circled in blue and orange, respectively.

We find that $V_O$ is non-magnetic and can be easily accommodated by the lattice without significant structural distortion (see Fig. 1(b)). This is unlike the case of the tin vacancy ($V_{Sn}$) as we show in Fig. S2(a) in the SM wherein O atoms immediately surrounding $V_{Sn}$ are displaced away from the defect site by as much as 0.3 Å. Furthermore, $V_{Sn}$ is spin-polarized with a net magnetic moment of 2 $\mu_B$ which mainly arises from the *p*-orbital contributions of Sn and O atoms located close to the vacancy site [33]. The calculated formation energies $E_f$ of $V_O$ and $V_{Sn}$ are 1.72 eV (4.15 eV) and 3.35 eV (0.92 eV) respectively under Sn-rich (O-rich) conditions. As the defect population is exponentially related to $E_f$ according to the Boltzmann distribution, growth conditions can ultimately influence the concentration of $V_{O/Sn}$ in the system. The competition between native vacancy states in monolayer SnO can thus be summarized as follows: in Sn-rich (or equivalently, O-poor) conditions, $V_O$ formation is strongly favored due to its low $E_f$. However, by switching to an O-rich (or Sn-poor) growth environment, for example by increasing the partial pressure of $O_2$, $V_{Sn}$ becomes the dominant defect and the system transitions towards a magnetic state. Therefore, Sn-rich conditions are an important prerequisite for facilitating $V_O$ formation and hence substitutional activity at the O site of monolayer SnO.

Next, the diffusive kinetics of O vacancies in SnO is investigated. With reference to the defect structure in Fig. 1(b), two plausible mechanisms for $V_O$ diffusion are identified, involving the migration of either (1) the first-nearest O neighbor to the defect site ($1NN_O$, circled blue), or (2) the second-nearest O neighbor to the defect site ($2NN_O$, circled orange). The structural evolution for both diffusion paths are shown in Figs. 2(a)-(b), and their corresponding energy profiles are given in Fig. 2(c). We anticipate the dynamic behavior of $V_O$ to be primarily driven by $1NN_O$ migration owing to its relatively low diffusion barrier $E_d$ of 1.42 eV as opposed to that of $2NN_O$ migration (1.97 eV). The higher barrier for $2NN_O$ diffusion may be linked to the indirect route taken by the O atom whereby it is temporarily dislodged from the O plane and inserted as an interstitial into the surface Sn plane, as shown in the transition state (TS) in Fig. 2(b).

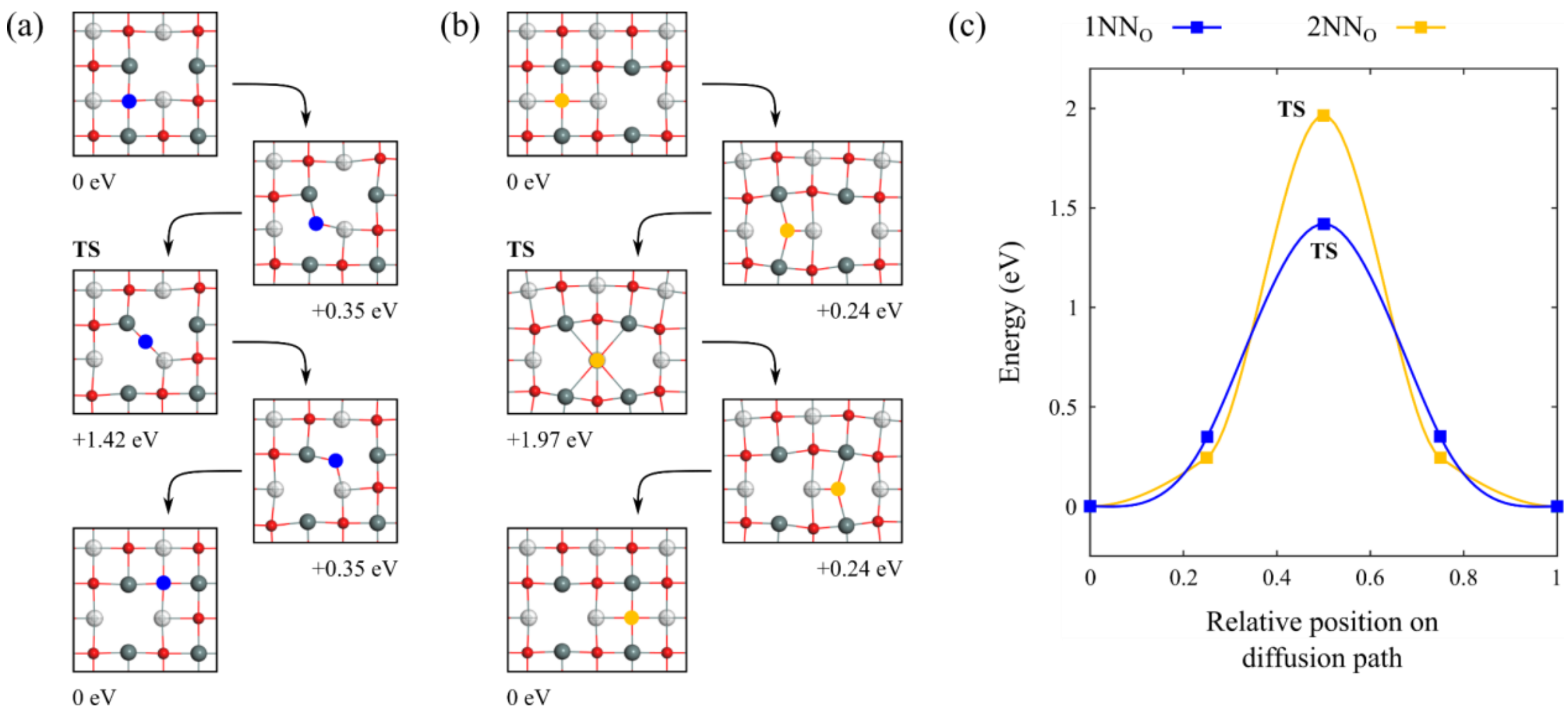

**Figure 2.** Snapshots of the local structure (top view) during O vacancy diffusion via (a) $1NN_O$ (blue), and (b) $2NN_O$ (orange) migration. (c) CINEB-calculated energy profiles for the $1NN_O$ and $2NN_O$ diffusion paths.

For a more contextual assessment, our results on the formation energy and diffusion barrier of both $V_O$ and $V_{Sn}$ in monolayer SnO are collated against available theoretical data for other common 2D (monolayer) materials [44-46], namely $MoS_2$ ($V_S$), h-BN ($V_B$, $V_N$), graphene and phosphorene, as shown in Fig. 3. Here, the upper/lower limits of $E_f$ for heteropolar materials (indicated by a vertical line) correspond to elemental rich/poor conditions. The $E_d$ of $V_{Sn}$ is deduced as 0.89 eV following a similar analysis of the nearest-neighbor diffusion pathways, as detailed in Fig. S2 in the SM. We note that graphene and h-BN, which are planar materials bound by strong C-C and B-N covalent bonds respectively, tend to exhibit very high $E_f$ values in excess of ~5 eV. In comparison, SnO possesses a relatively low to moderate $E_f$ for both $V_O$ (1.72 – 4.15 eV) and $V_{Sn}$ (0.92 – 3.35 eV) due to its softer and more flexible Sn-O bonds [40]. These values are typical of non-planar (buckled/stacked) structures and are observed to be of similar range to the $E_f$ of phosphorene vacancies (1.65 eV) and $V_S$ in $MoS_2$ (1.22 – 2.25 eV).

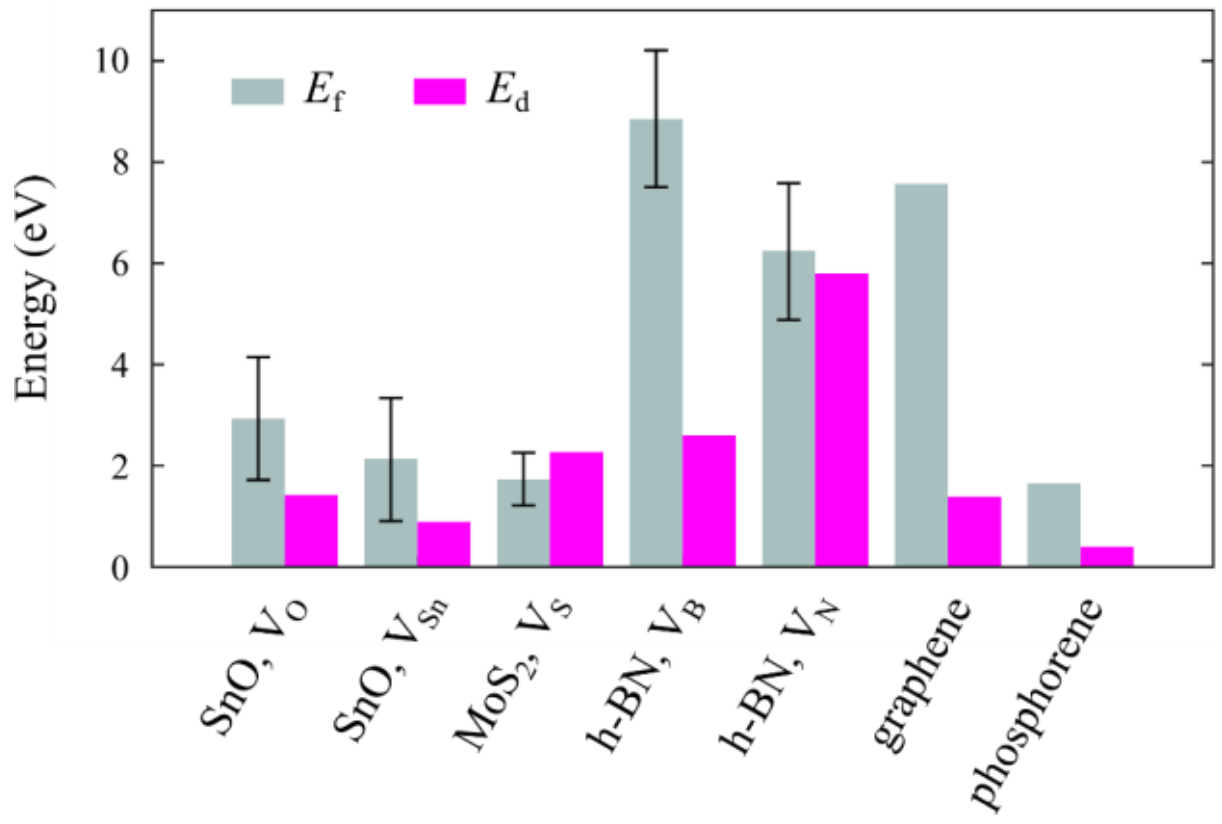

**Figure 3.** Formation energies and diffusion barriers of vacancy defects across various 2D (monolayer) materials. For heteropolar materials, the uncertainty in formation energy due to the variation in chemical potential is indicated by a vertical line. The relevant data for $MoS_2$, h-BN, graphene and phosphorene are adapted from available literature [44-46].

The diffusion barrier $E_d$ of a vacancy is reflective of its mobility according to the Arrhenius relation which defines the vacancy hopping frequency as $v = v_s e^{-E_d/k_B T}$, where $v_s$ is a characteristic prefactor given by the Vineyard formula [47] (usually ~$10^{13}$ Hz), $k_B$ is the Boltzmann constant and $T$ is the temperature. More simply, this means that a lower (higher) $E_d$ ($T$) will increase vacancy mobility $v$. Among the 2D materials presented in Fig. 3, phosphorene vacancies have the smallest $E_d$ (0.40 eV) and are thus extremely mobile, even at room temperature [44]. Although the native vacancies in SnO have higher $E_d$ values ($V_O$: 1.42 eV, $V_{Sn}$: 0.89 eV), they can still be fairly mobile, especially under elevated temperatures. Our calculated value of $E_d$ for $V_O$ is further noted to be highly comparable to that of graphene vacancies (1.39 eV) for which active diffusion has been experimentally observed using transmission electron microscopy [48]. The itinerant behavior of $V_O$, in addition to its low $E_f$ in Sn-rich conditions, render O vacancies in 2D tin monoxide worthwhile of deeper assessment. As we will reveal later in this work, these defects can offer fertile ground for localized substitutional doping and the tuning of electronic properties, thereby opening up new opportunities for atomically thin SnO in nanoscale applications.

### 3.2. Substitutional doping at the O site

Having shown that O vacancies can be formed in monolayer SnO, predominantly under Sn-rich conditions, we proceed to study how they can be consecutively filled by a dopant. Here, we consider various substitutional dopants X = C, Si, N, P, S, F, Cl, H, $H_2$ occupying the O site, as illustrated schematically in the top panel of Fig. 4(a). These dopants may be classified as double acceptors (C, Si), acceptors (N, P), isoelectronic species (S), donors (F, Cl) or hydrogen species (H, $H_2$). Unless otherwise stated, all findings presented hereon are for the general case of isolated doping in Sn-rich conditions.

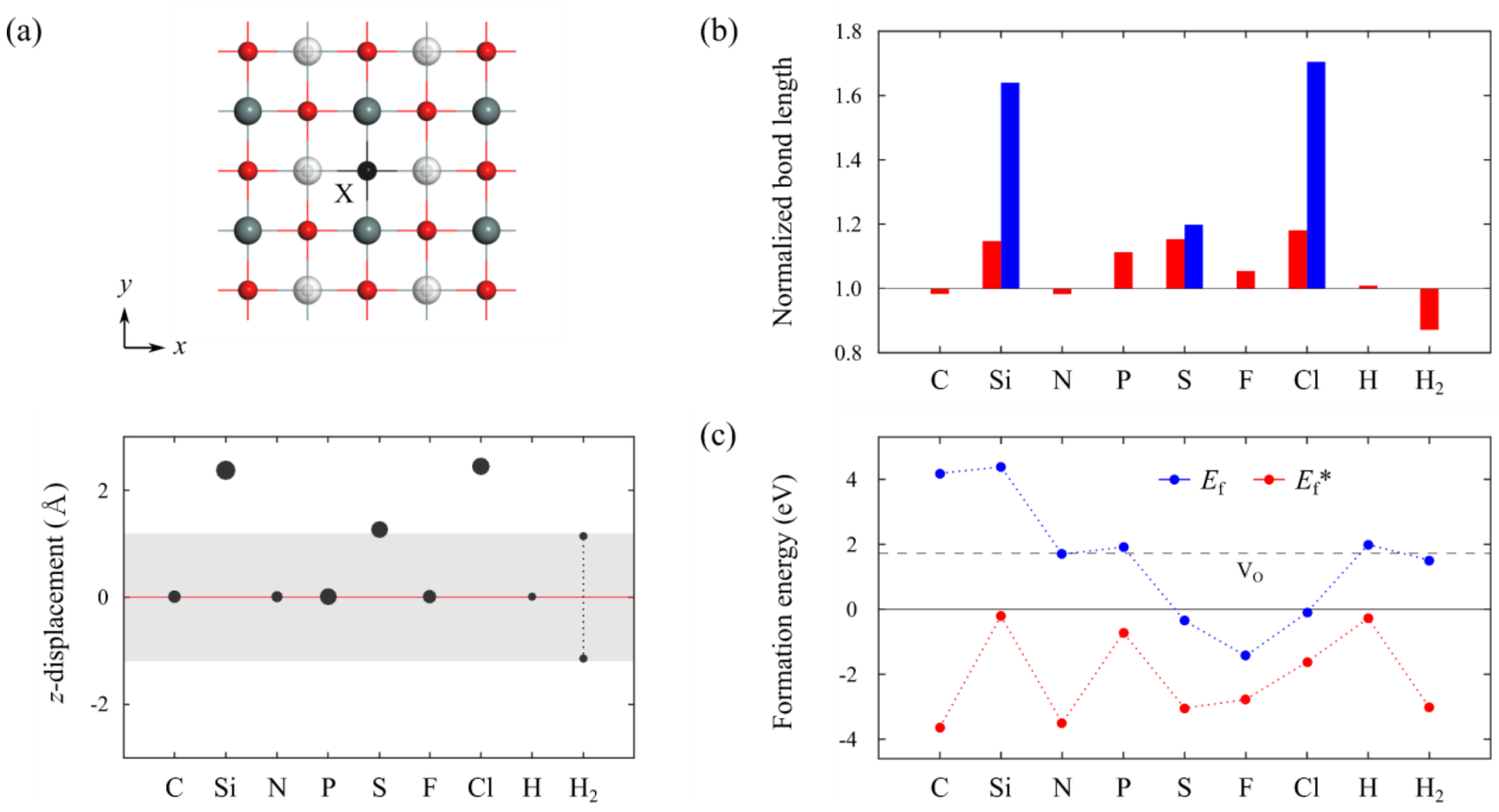

**Figure 4.** (a) Top panel: Atomic model (top view) of a substitutional dopant (X = C, Si, N, P, S, F, Cl, H, $H_2$) at the O site of monolayer SnO. Bottom panel: Vertical ($z$) displacement of the dopants from the O site after full structural optimization, with markers scaled according to the dopants' atomic radii. The monolayer region is shaded grey. (b) Sn-X bond lengths normalized with respect to the equilibrium bond length of Sn-O. For the adsorbed species Si, S and Cl, red and blue bars refer to the bonds formed with the near and far surfaces, respectively. (c) Formation energies of X-doped monolayer SnO based on the bulk/molecular (blue) and atomic (red) chemical potentials of the dopant. Note that the formation energy for $H_2$ substitution is given on a per molecule basis.

The results of our analysis on the structures of X-doped monolayer SnO are provided in the bottom panel of Fig. 4(a) and in Fig. 4(b). Notably, not all dopants achieve perfect O-site substitution. From Fig. 4(a) (bottom panel) which shows the vertical displacement of the dopants from the O site, it is apparent that only C, N, P, F, H and $H_2$ species can be perfectly substituted into the lattice. The $H_2$ molecule in particular undergoes dissociation about the O site and each H atom takes up an equivalent position on either surface Sn plane. For brevity, we refer to this group of dopants as $X_P$. The remaining investigated species (i.e., Si, S, and Cl) – which we categorize as $X_{ad}$ – are found to be too large to be accommodated within the SnO monolayer and are instead adsorbed just above the surface as $X_{ad}$-$V_O$ type complexes. Atomic models of all the X-doped structures are supplemented in Fig. S3 in the SM. We have also calculated the planar-averaged electrostatic potential along the normal (*z*) direction for all systems in Fig. S4 in the SM. Our results show that the vacuum level is identical on both sides of the surface for all structures, including those containing adsorbed dopants on one side, verifying that the prescribed vacuum region of ~20 Å is sufficient for our slab models and no spurious dipole moments are present.

The $X_P$ dopants form four major Sn-X bonds akin to pristine SnO, whereas the $X_{ad}$ dopants form two major bonds to the near surface and two weaker, long-range bonds to the far surface of the monolayer accordingly. The relevant bond lengths for all dopants are normalized with respect to the equilibrium bond length of Sn-O (2.26 Å) and given in Fig. 4(b). The adsorbed species Si, S and Cl are larger compared to the other dopants and their bonds with Sn are evidently much longer, where Sn-Si = 2.59 Å, Sn-S = 2.61 Å and Sn-Cl = 2.67 Å (see red bars). This thus prevents these species from being fully assimilated within the monolayer. Among the $X_P$ dopants, C, N, F and H species bear rather comparable bond lengths (2.22 – 2.38 Å) to that of Sn-O, indicative of their superior compatibility with the host lattice. On the contrary, the P dopant forms relatively long bonds (2.51 Å) within the monolayer. This can propagate extensive lattice distortion and compromise the structural integrity of SnO in practice. Furthermore, between H and $H_2$ dopants, the latter appears to yield a more stable state of passivation due to its much shorter Sn-H bonds (1.97 Å). Of the $X_{ad}$ dopants, we find that Si and Cl barely interact with the far surface (bond lengths > 3.6 Å), essentially leaving the Sn atoms on this side unsaturated and potentially open to functionalization by a second dopant atom.

The formation energies $E_f$ and $E_f$* of the X-doped structures are charted in Fig. 4(c) based on the bulk/molecular (blue) and atomic (red) chemical potentials of the dopant, respectively. All substitutions are clearly spontaneous ($E_f^* < 0$) with respect to the free dopant atom. However, if we consider more realistic chemical potentials of the dopants in their more stable bulk or molecular form, this trend no longer holds true. For instance, due to the extremely stable diamond structures of bulk C and Si, both C and Si substitutions are quickly understood to be highly unfavorable, with $E_f > 4$ eV. As for the molecular dopants, only F doping is highly spontaneous, S and Cl substitutions are marginally favored, while the remaining dopants have positive $E_f$ values which are comparable to that of $V_O$ itself (see dashed line). Confirming our earlier suspicion, the $H_2$ dopant, or 2H as it becomes upon dissociation, is energetically preferred at the O site over just a single H atom. The calculated $E_f$ is always higher than $E_f$* since $E_f$ takes into account the energy cost of dissociating the $X_2$ molecule. For example, in N

(F) doping, this effect can be very significant (slight) because the $N_2$ ($F_2$) molecule has a relatively high (low) binding strength of 10.4 eV (2.7 eV) – taken from Table S1 in the SM. Despite having positive formation energies that do not favor doping in equilibrium conditions, substitution may still be achieved under external stimuli by promoting the dissociation of dopant molecules catalytically [49] or through electron beam irradiation as previously demonstrated in boron-nitride nanostructures [50, 51] and 2D transition metal dichalcogenides [52]. Hence, postsynthesis electron-mediated doping of the O site could be a possible strategy for tailoring the structure and properties of ultrathin SnO sheets.

The presence of impurities such as vacancies and dopants will inevitably introduce residual strain fields in the SnO lattice. To shed light on the mechanical distortion induced by the various defects (V or X) in the monolayer, the strain distribution of the surrounding Sn-O bonds is analyzed. As we illustrate in Fig. 5(a), the local environment around V or X can be segmented into three main regions for visual clarity: I ($0 < d \leq a$), II ($a < d \leq 2a$), and III ($d > 2a$), where $a$ is the lattice constant and $d$ is the distance between an arbitrary Sn-O bond and the defect site. By indexing the Sn-O bonds in these three regions in order of increasing $d$, the variation in strain as one moves away from the defect site can be mapped. The strain fields associated with the vacancy defects $V_O$ and $V_{Sn}$ are shown in Fig. 5(b), while those of the O-site dopants are given in Fig. 5(c).

As expected, the strain field produced by the O vacancy is far less pronounced and not as pervasive as that around the Sn vacancy. In general, the impurities $V_O$, $V_{Sn}$ and $X_P$ generate isotropic residual strain which decays very smoothly and quickly into the lattice, however the strain fields around $X_{ad}$-$V_O$ complexes are inherently polarized by one-sided functionalization, giving rise to highly chattered, long-range signatures. To this end, additional pinning of $V_O$ by a second dopant atom on the other side of the monolayer may serve to improve the isotropy as well as decay rate of the strain field. The defect cores of the isotropic impurities (see red arrows) can be positive (C, N, P, F) or negative ($V_O$, $V_{Sn}$, H, $H_2$), reflecting the tensile or compressive nature of the nearest Sn-O bonds to the defect, while its magnitude indicates the extent of structural mismatch between the defective and pristine condition. This might suggest that *externally applied* tension (compression) could play a critical role in stabilizing the positive (negative) defect core, in a bid to neutralize strain field fluctuations and promote defect stability. Furthermore, while not studied, the stable coexistence of adjacent vacancies and dopants, including various co-doping possibilities could also perhaps be driven under dipole-like conditions involving isotropic impurities of oppositely cored polarities.

Among the different dopant species studied, we find that N and F can be immersed almost seamlessly within the SnO lattice, rendering them as ideal candidates for O-site substitution with little detriment to the mechanical integrity of the structure. The desirability of N doping has been corroborated in a recent study by Becker *et al.* [53] where nitrogen-doped SnO thin films with *p*-type conductivity were shown to display exceptional long-term stability over a period of four months. In contrast, the researchers reported rapid outdiffusion and degradation of hydrogen-doped samples over the same time duration, which may be ascribed, at least in part, to differences in mechanical compatibility. To conclude this section, we refer the reader to Table S2 in the SM for a full summary of our characterization results for all the investigated impurities.

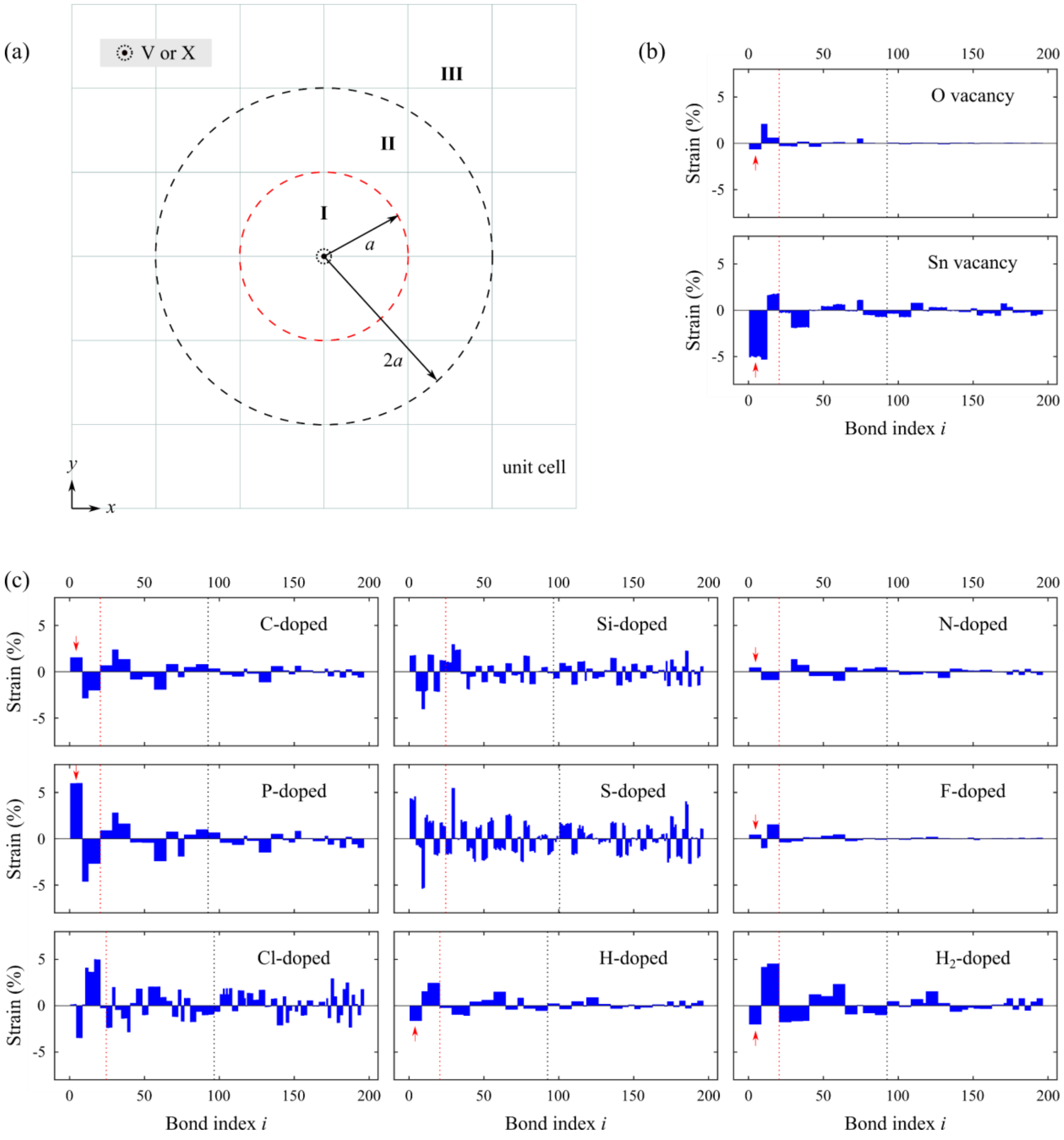

**Figure 5.** (a) Schematic diagram of the local environment around the V or X impurity (top view) showing the three main regions I, II and III. Strain fields associated with the (b) vacancy defects ($V_O$, $V_{Sn}$), and (c) O-site dopants (X = C, Si, N, P, S, F, Cl, H, $H_2$). The upper boundaries of regions I and II are indicated by red and black dotted lines, respectively.

### 3.3. Electronic properties of doped monolayer SnO

In this section, the electronic implications of doping at the O site, including the effect of dopant concentration, is examined in monolayer SnO. Firstly, from the local density of states (LDOS) of the O-site dopants (see Fig. 6), we verify that N and P can behave as *p*-type acceptors in the substitutional system with gap states introduced near the valence band maximum (VBM), whereas F can act as an *n*-type donor given the presence of gap states near the conduction band minimum (CBM). Conversely, the double acceptors (C, Si) and $H_2$ have discrete energy levels that lie toward the middle of the gap, while S, Cl and H do not generate any significant LDOS in the bandgap at all. The orbitals of N, P, F, Cl and H, which are spin-polarized, contribute to total net magnetic moments of 1 $\mu_B$ observed in these systems.

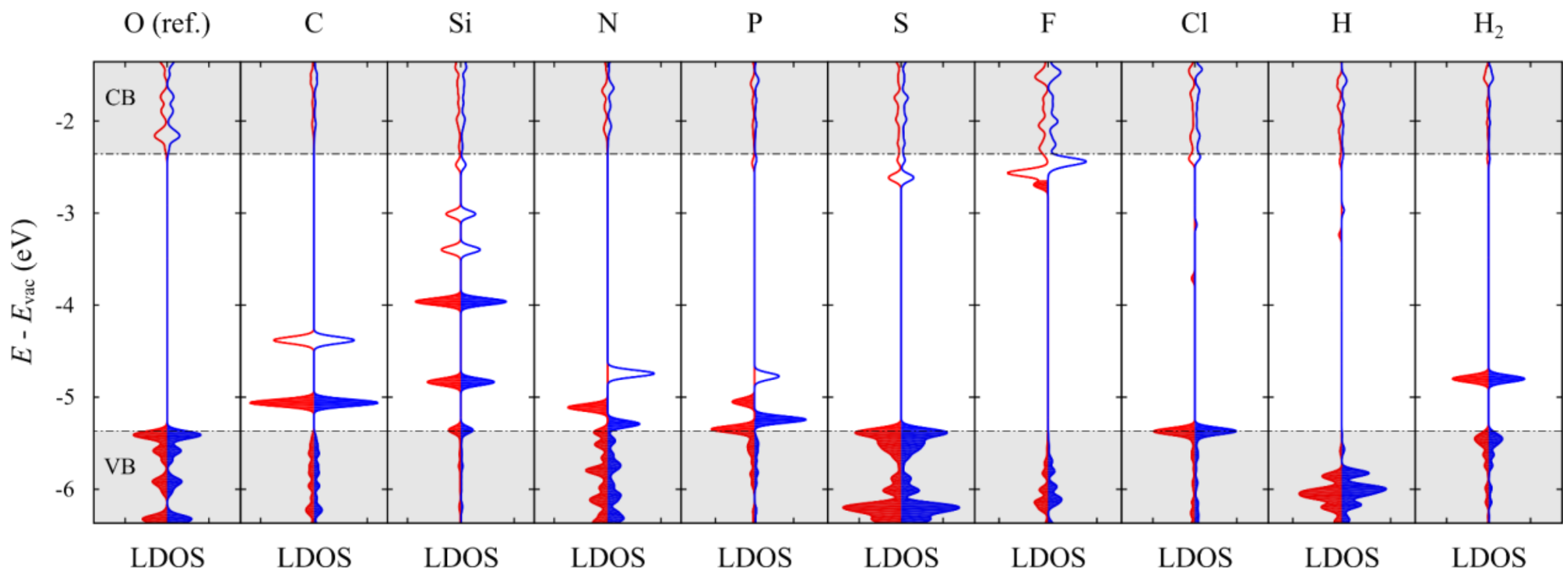

**Figure 6.** LDOS of the O-site dopants, aligned by setting the vacuum level $E_{vac}$ of each system at zero. The reference LDOS of O is given in the left-most panel. Spin-up and spin-down components are indicated in red and blue accordingly, while the grey-shaded regions denote the conduction band (CB) and valence band (VB) edges of the pristine monolayer. Occupied (unoccupied) states are represented by filled (empty) regions in the plots.

Using Bader charge analysis [54], we are also able to quantify how much charge is transferred to the dopant during functionalization (see Table S2 in the SM). Negative values in all cases simply mean that in forming Sn-X bonds, Sn will lose some of its valence electrons to the dopant just as it does to surrounding O atoms. The net charge transfer in the pristine monolayer is calculated to be −1.25*e*/O atom. In comparison, the amount of charge transferred to the dopant is found to be slightly higher for C (−1.30*e*) and N (−1.41*e*), yet considerably lower for the other dopants. Notably, the charge transfer can be as low as −0.54*e* for H and −0.46*e* for Si among the $X_P$ and $X_{ad}$ group of dopants, respectively.

The effects of high-concentration doping are evaluated for the perfect substitutionals $X_P$ = C, N, P, F, H, $H_2$ by simulating the $SnO_{(1-\delta)}X_{\delta}$ system in the degenerate limit $\delta = 0.5$. At this concentration level, which is 25 times higher than what we have considered all along in the case of isolated doping ($\delta = 0.02$), interdopant coupling becomes significant. Here, a unit cell model is adopted to obtain a rudimentary benchmark for isotropic doping at the critical upper limit where the O:X ratio is 1:1. From a theoretical standpoint, an understanding of this configuration can be useful at the dopant screening stage as it may provide an early indication of the sort of properties one may come to expect from a particular candidate. More exhaustive sampling of the potential energy surface across intermediate concentrations and various dopant distribution profiles is nonetheless further warranted in future works to bridge the gap between simulation and experiments.

The optimized lattice constants *a* and formation energies $E_f$ per formula unit (f.u.) of the various degenerately doped phases are reported in Fig. 7(a) and Table 1. Monolayers hybridized with C, N and P exhibit large lattice constants relative to pristine SnO (see dashed line; $a^{SnO}$ = 3.83 Å), while those containing F, H and $H_2$ have smaller values. It is noted that the positive (negative) defect cores identified previously for the isolated dopant (see Fig. 5(c) and Table S2 in the SM) can be fairly predictive of the resulting structural expansion (shrinkage) under heavy doping. Apart from $SnO_{0.5}H_{0.5}$ having a net magnetic moment of 0.15 $\mu_B$, the other investigated systems are found to be non-magnetic.

**Table 1.** Calculated net magnetic moment *m*, geometric parameters, and formation energy $E_f$ per formula unit (f.u.) of monolayer $SnO_{(1-\delta)}X_\delta$ in the degenerate limit $\delta$ = 0.5.

| Species, X | Chemical formula | *m* ($\mu_B$) | Geometric parameters (Å) | | $E_f$ per f.u. (eV) |
|---|---|---|---|---|---|
| | | | *a* | *t* | |
| C | $SnO_{0.5}C_{0.5}$ | 0.00 | 4.066 | 1.935 | 0.566 |
| N | $SnO_{0.5}N_{0.5}$ | 0.00 | 3.950 | 2.116 | −0.463 |
| P | $SnO_{0.5}P_{0.5}$ | 0.00 | 4.376 | 2.411 | −0.513 |
| F | $SnO_{0.5}F_{0.5}$ | 0.00 | 3.706 | 2.823 | −2.138 |
| H | $SnO_{0.5}H_{0.5}$ | 0.15 | 3.566 | 2.754 | −0.384 |
| $H_2$ | $SnO_{0.5}H$ | 0.00 | 3.692 | 2.880 | −0.187 |

The formation of all degenerately doped compounds, with the exception of $SnO_{0.5}C_{0.5}$, appear to be energetically favorable ($E_f$ < 0), in support of their synthesis and isolation experimentally. More interestingly, $SnO_{0.5}F_{0.5}$ (−2.14 eV) possesses a remarkably low $E_f$ value that is very close to that of pristine SnO (see dashed line; $E_f^{SnO}$ = −2.44 eV), implying superior oxidation resistance and stability of this phase in air once formed. The remaining compounds $SnO_{0.5}N_{0.5}$, $SnO_{0.5}P_{0.5}$, $SnO_{0.5}H_{0.5}$ and $SnO_{0.5}H$ could nevertheless still be of practical value, but they may require stringent environment control and/or suitable passivation techniques (e.g., capping by overlayers) in order to sustain long-term stability and function. Our charge transfer results for the degenerately doped monolayers (see Fig. 7(b)) remain in good agreement with earlier analysis for the isolated dopant, with reduced Sn-X charge transfer capacities clearly reflected in the case of P, F, H and $H_2$ functionalization.

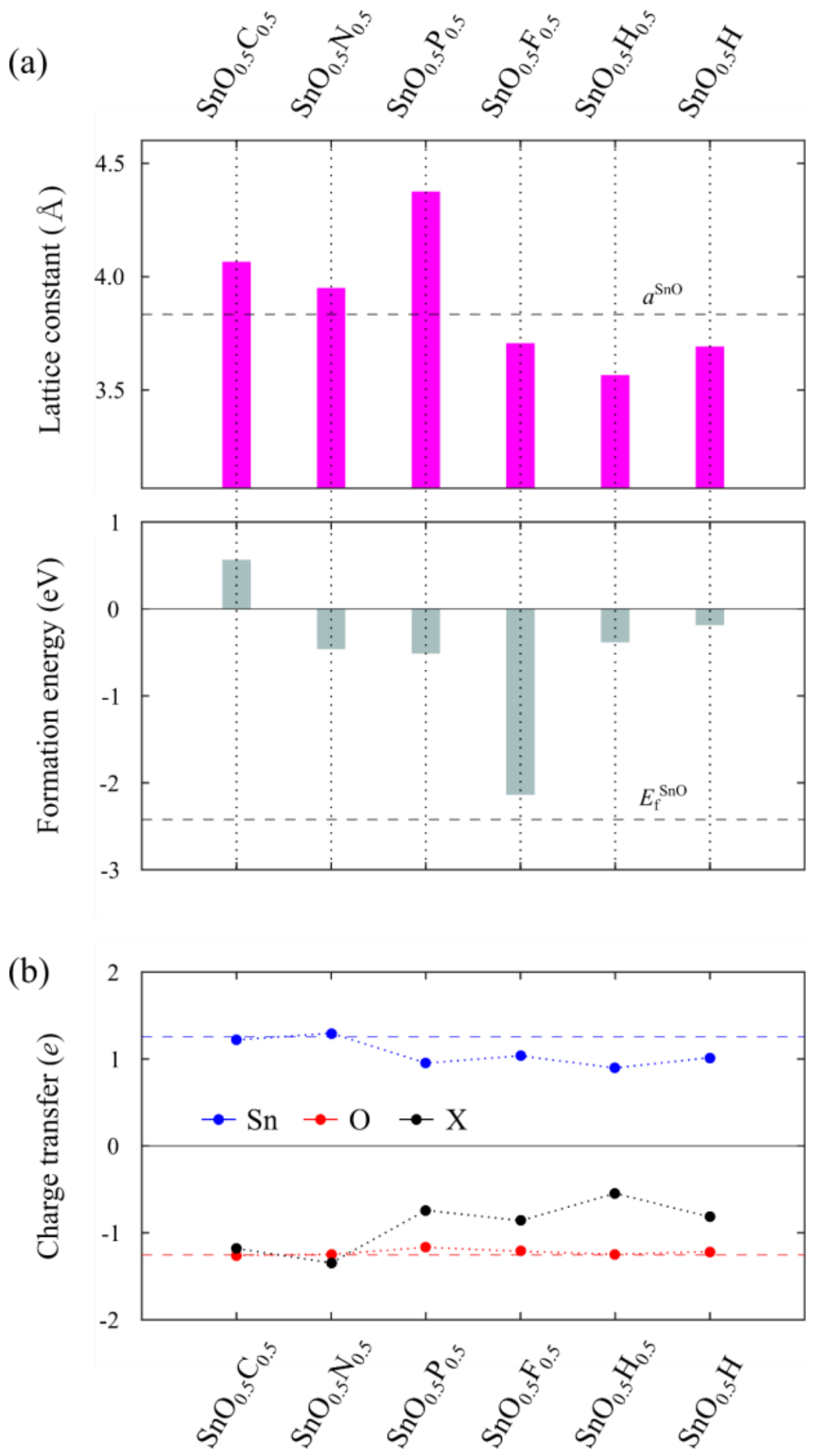

**Figure 7.** (a) Lattice constants and formation energies of $SnO_{(1-\delta)}X_{\delta}$ monolayers (X = C, N, P,

F, H, $H_2$) in the degenerate limit $\delta = 0.5$. (b) Net charge transfer among Sn, O and X constituents of the various phases. The reference charge transfer of ±1.25*e* in pristine monolayer SnO is indicated by blue (Sn) and red (O) dashed lines.

The influence of O-site doping on the work function of monolayer SnO is illustrated in Fig. 8(a). Evidently, relative to the pristine condition (see dashed line; $WF^{SnO}$ = 5.17 eV), the work function can be lowered and tuned over a broad range of 2.64 − 4.93 eV depending on the choice of substitutional species and its concentration. For acceptors such as C, N and P, the work function generally lies in the upper range and tends to decrease with increasing dopant concentration, while the reverse is true for donors like F. This is because, as the concentration is increased, impurity atoms are more densely packed across the lattice thus strengthening the electronic interactions that take place between them. Consequently, more and more discrete impurity levels are introduced near the VBM (acceptor) or CBM (donor), which will gradually broaden and merge together with the valence or conduction bands, thereby resulting in new band edges with a higher or lower Fermi level accordingly, that is, with respect to the vacuum level. This mechanism is corroborated by analysis of the density of states (DOS), given in Fig. S6 in the SM, which shows the individual contributions to the DOS by the parent SnO compound and dopant species at both isolated and degenerate doping conditions. On this basis, our results assert that H and $H_2$ will behave as an *n*-type donor and *p*-type acceptor respectively when incorporated at the O site. Their substitution energetics (see Fig. 4(c)) however reveal that vacancy filling by $H_2$ offers a more stable doping channel, which seems to agree well with experimental observations of *p*-type conductivity in hydrogen-doped SnO thin films [53].

The electronic band structures of the degenerately doped phases are supplemented in Fig. 8(b) and Fig. S5 in the SM. As a reference, the pristine SnO monolayer is noted to have a rather wide indirect bandgap of 3.01 eV, as shown in Fig. 8(b) (left panel). In addition to impurity-induced distortions in the band structure, we document the metallization of almost all systems under heavy doping. Only the $SnO_{0.5}H$ monolayer continues to remain a semiconductor, albeit with a substantially reduced bandgap of ~0.2 eV.

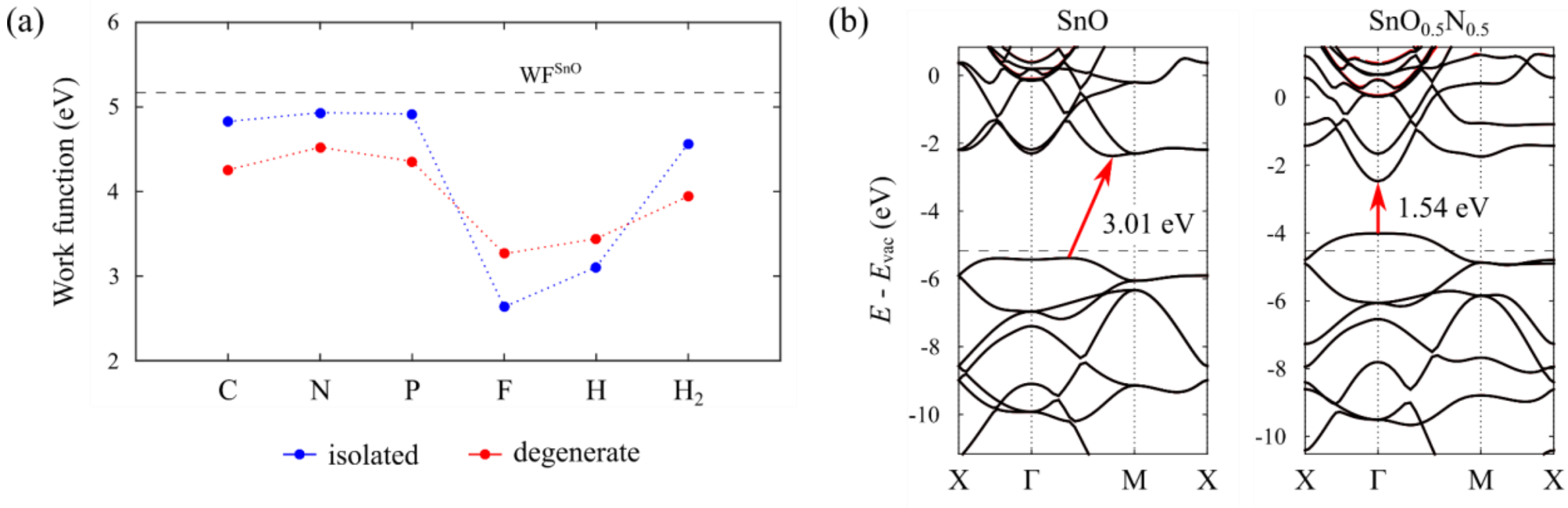

**Figure 8.** (a) Work function of $SnO_{(1-\delta)}X_{\delta}$ monolayers (X = C, N, P, F, H, $H_2$) in the isolated ($\delta$ = 0.02, blue) and degenerate ($\delta$ = 0.5, red) limits. (b) Electronic band structures of SnO (pristine) and $SnO_{0.5}N_{0.5}$ monolayers, decomposed into spin-up (red) and spin-down (black) band components. The vacuum level $E_{vac}$ is set at zero and the Fermi level is indicated by a dashed line.

In the $SnO_{0.5}N_{0.5}$ monolayer (see Fig. 8(b) (right panel)), an appreciable energy interval of 1.54 eV is elucidated above the Fermi level within which no electron states are found. Intriguingly, the upper and lower edges of this interval both lie at the Γ point, resembling that of a direct bandgap; no such feature is found in all other cases. The activation of an indirect-to-direct bandgap transition is unprecedented in layered tin monoxide, making it especially important if one can be realized through N doping. Although the $SnO_{0.5}N_{0.5}$ phase reported here is predicted to be metallic, the lower edge of this interval remains very close to the Fermi level ($\Delta E$ ~0.5 eV). This suggests that semiconducting properties may be achieved with appropriate tailoring of the concentration and/or distribution of nitrogen dopants in the system. Besides, mechanical deformation by strain or pressure has also been shown to induce metal-to-semiconductor switching in a variety of materials [55-57] and can thus offer a complementary means for promoting electron reorganization. This unique finding highlights exciting opportunities in the Sn-O-N ternary system and may lead to useful applications in photocatalysis and high-performance optoelectronics.

**4. Conclusions**

In this work, we first characterized the structural dynamics of the oxygen vacancy defect ($V_O$) in monolayer SnO, revealing key insights into its formation energetics and mechanisms of diffusion within the material. Subsequently, we addressed the possibility of doping at the O site by complementary lightweight species, namely C, Si, N, P, S, F, Cl, H and $H_2$. Overall, it appears that a feasible strategy for O-site doping may be that of growing O-poor samples of SnO, followed by filling of the vacancies with an appropriate dopant as desired. From both a mechanical and electronic perspective, N and F dopants are identified as ideal candidates for substitution, endowing monolayer SnO with *p*-type and *n*-type conductivities, respectively. In the limits of heavy doping, we postulate $SnO_{0.5}F_{0.5}$ to be highly air-stable, and the $SnO_{0.5}N_{0.5}$ phase to exhibit a “direct-gap” configuration that may be of functional value in photonics and other light-based applications. Extensions of this work in the computational domain will include employing high-throughput screening based on machine learning and/or cluster expansion techniques to precisely investigate the relationship between dopant concentration, structure and electronic properties. This study also advocates further research on defects in 2D layered tin monoxide, particularly in relation to the role of charged states and the aggregation of higher-order defects such as co-dopant states and substitutional-vacancy complexes. The findings reported here cater to both theoretical and experimental interests to develop targeted doping strategies and produce hybridized nanosheets of SnO with unusual chemical, optical and electronic properties.

## Acknowledgments

This research article was supported by the Economic Development Board, Singapore and Infineon Technologies Asia Pacific Pte. Ltd. through the Industrial Postgraduate Programme with Nanyang Technological University, Singapore, and the Ministry of Education, Singapore (Academic Research Fund TIER 1-RG174/15). The computational work for this article was partially performed on resources of the National Supercomputing Centre, Singapore (https://www.nscc.sg).

## Supplementary material available

The supplementary material contains supporting data on the reference phases adopted in this work, a summary of our defect characterization results, and DFT simulation results (diffusion analysis, structural configurations, planar-averaged electrostatic potential plots and electronic band structures) accessory to the main article. This material is available free of charge via the Internet at [*insert article doi here*].